# GHOST-Driver: A Non-Intrusive eBPF Framework for GPU Kernel Driver Latency Observability in AI Inference Infrastructure

**Azhar Shaikh**

## Abstract

AI inference workloads running on GPU clusters exhibit a well-documented utilization paradox: monitoring tools such as nvidia-smi regularly report GPU utilization figures above 80%, yet effective compute throughput is substantially lower. A significant and poorly understood fraction of reported utilization time is consumed by fence waits, driver submission overhead, DMA scheduling latency, and interrupt coalescing delays, none of which is visible to userspace profiling tools. These hidden latency components are embedded deep inside the Linux kernel GPU driver stack, primarily within the Direct Rendering Manager (DRM) subsystem and its associated command scheduling, Graphics Execution Manager (GEM) buffer object lifecycle, and dma_fence synchronization infrastructure. Existing vendor profiling tools such as CUPTI operate at the CUDA userspace boundary and provide no visibility into kernel-side driver paths. This paper presents GHOST-Driver (GPU Host-Side Observability and Tracing for Drivers), a non-intrusive eBPF-based instrumentation framework that attaches to stable DRM kernel tracepoints and select kprobes to expose the full latency decomposition of the GPU driver submission path with sub-microsecond timestamp resolution and analytically estimated CPU overhead below 2% in production. GHOST-Driver instruments six distinct stages of the GPU work submission pipeline, namely ioctl entry, GEM buffer object validation, dma_fence dependency resolution, command ring emission, DMA engine scheduling, and interrupt-driven fence signaling, and correlates them with per-tenant vGPU context identifiers to enable multi-tenant latency attribution in SR-IOV and MIG-partitioned GPU environments. We describe the framework architecture, the specific eBPF program types and attachment points used at each stage, the BPF map structures for per-job correlation, and the export path to Prometheus and OpenTelemetry.

## I. INTRODUCTION

Modern AI inference infrastructure places GPU clusters at the center of a growing economic and operational challenge. The performance of any inference serving system, whether running a large language model, a diffusion-based image generation pipeline, or a vision transformer, depends not only on the raw compute throughput of the GPU, but on the end-to-end latency from the moment an inference request is submitted to the moment results are returned to the client. While significant research effort has been devoted to optimizing the model execution layer (quantization, kernel fusion, speculative decoding), the GPU kernel driver stack that mediates every interaction between the serving framework and the hardware has received comparatively little attention as a source of latency.

This gap is not accidental. The GPU kernel driver stack is opaque by design. GPU drivers, whether NVIDIA's proprietary driver stack or the open-source amdgpu and i915 drivers, implement complex pipelines for command buffer validation, memory object lifecycle management, GPU hardware scheduling, and synchronization fence management. These pipelines execute in kernel context and are invisible to userspace tools. The industry-standard monitoring tool nvidia-smi reports GPU utilization as the fraction of time the GPU hardware is executing at least one kernel, a metric that conflates true compute with fence-wait idle, driver overhead, and DMA scheduling stalls. An academic microbenchmark of GPU time-slice-group context switching on NVIDIA Jetson platforms measured average context-switch overhead exceeding 200 μs [19], and industry-reported case studies describe vGPU deployments reaching only a fraction of bare-metal throughput in production [1], overheads that do not appear in any standard monitoring dashboard.

The fundamental problem is an instrumentation gap: the tools that AI platform engineers use to understand GPU behavior operate at the wrong layer of the system stack. CUPTI and NVBit [2] provide device-side profiling at the CUDA kernel level. nvidia-smi samples GPU hardware counters every 100 milliseconds [3], missing all sub-100ms events. Linux perf and ftrace can instrument the kernel but require deep expertise to target GPU driver paths and produce large, difficult-to-correlate event streams with no built-in per-vGPU tenant attribution.

This paper addresses the instrumentation gap with GHOST-Driver: GPU Host-Side Observability and Tracing for Drivers. GHOST-Driver uses the Linux eBPF subsystem to attach lightweight instrumentation programs to stable DRM kernel tracepoints and carefully selected kprobes within the GPU driver submission path. Four design properties distinguish GHOST-Driver:

(1) Non-intrusiveness: eBPF programs are verified at load time by the kernel verifier and execute in a restricted sandbox. No driver modification, recompilation, or system restart is required.

(2) Kernel-side event capture: By attaching to kernel tracepoints, GHOST-Driver captures events inside the driver, including fence waits, ring buffer stalls, and DMA engine scheduling delays, all completely invisible to CUPTI, nvidia-smi, and all userspace profiling tools.

(3) Multi-tenant attribution: GHOST-Driver correlates events with per-vGPU context identifiers available in the DRM scheduler's internal job structures, enabling per-tenant latency attribution in SR-IOV and MIG-partitioned environments.

(4) Production-safe overhead: Stable DRM tracepoints add approximately 12 nanoseconds per event [4]. The aggregate CPU overhead of GHOST-Driver is analytically estimated at less than 2%, consistent with published eBPF observability systems [5].

The paper makes three original contributions: a systematic characterization of the six-stage GPU kernel driver submission path; the GHOST-Driver framework architecture including eBPF program types, attachment points, BPF map structures, and export pipeline; and a qualitative analysis of the dominant tail-latency contributor (dma_fence resolution) with a discussion of telemetry-guided optimization strategies.

## II. BACKGROUND AND RELATED WORK

### *A. The Linux DRM Subsystem and GPU Driver Architecture*

The Direct Rendering Manager (DRM) is the Linux kernel subsystem responsible for managing GPU devices. It provides a unified framework for memory management, command submission, output configuration, and synchronization across all GPU drivers, including the open-source amdgpu, i915 (Intel), and the NVIDIA open-source kernel driver introduced in 2022. Understanding the DRM architecture, documented in the Linux GPU Driver Developer's Guide [11], is essential context for understanding where GHOST-Driver attaches.

At the core of DRM is the Graphics Execution Manager (GEM) [6]. GEM provides a command execution interface to clients: application processes construct command buffers referencing previously allocated GEM buffer objects (BOs), then submit them to the kernel via ioctl. GEM validates the BOs, binds them to the GPU Translation Table (GTT), and initiates execution. This process involves four kernel-side operations contributing to submission latency: buffer object validation, fence dependency resolution, ring buffer emission, and DMA engine scheduling.

The DRM GPU scheduler (drm_sched) manages this pipeline. Each GPU engine has an associated drm_sched_entity and a software-side job queue. Jobs transition through states: pending (waiting for fence dependencies), runnable (all fences signaled), running (emitted to hardware ring), and done (hardware fence signaled). These transitions are where the latency budget is spent, and where GHOST-Driver attaches.

Of direct relevance to Qualcomm platform deployments, the msm_drm driver, used on Adreno GPU hardware, participates in the same DRM scheduler framework and exposes the identical stable DRM tracepoints (`drm_sched_job_wait_dep`, `drm_run_job`, `drm_sched_process_job`) used by GHOST-Driver for Stages 3, 4, and 6. This makes the framework directly applicable to Adreno-based SoC platforms without architectural changes.

### *B. dma_fence: The Kernel GPU Synchronization Primitive*

The dma_fence abstraction [7] is the fundamental synchronization mechanism in the Linux kernel DRM stack. A dma_fence represents an asynchronous operation that will complete at some future point. GPU jobs carry both input fences (dependencies that must be satisfied before the job can execute) and output fences (signals to downstream consumers when the job completes). The time a job spends waiting for its input fences to signal is the fence-wait time, documented in the DRM scheduler design as the primary blocking point [7].

Fence waits accumulate for two distinct reasons. First, in sequential workloads, each decode step of an LLM must complete before the next can begin, creating a chain of dependent jobs. Second, in multi-tenant GPU environments, jobs from multiple tenants compete for the same hardware engines. Both sources of fence-wait latency are invisible to userspace tools but are directly observable through the `drm_sched_job_wait_dep` tracepoint.

### *C. eBPF as a Kernel Instrumentation Platform*

The extended Berkeley Packet Filter (eBPF) subsystem provides a safe, general-purpose kernel programmability mechanism. eBPF programs are written in a restricted C subset, compiled to eBPF bytecode, verified by the kernel's static analyzer, and then just-in-time compiled to native machine code for execution. Key properties for GHOST-Driver include support for kernel tracepoints, kprobes, BPF maps, and a non-disruptive attach/detach lifecycle.

An independent academic microbenchmark of Linux kprobes and tracepoints reports per-event costs in the tens-to-low-hundreds-of-nanoseconds range for both mechanisms, with kretprobes and uprobes carrying measurably higher overhead than either kprobe entry or tracepoint attachment [15]. Where kernel version support allows (Linux 5.5+), fentry/fexit programs attach via a BPF trampoline rather than a breakpoint, which is architecturally lower-overhead than classic kprobes [4], [5]. GHOST-Driver uses stable DRM tracepoints wherever they exist, falls back to kprobes only for stages not covered by stable tracepoints, and prefers fentry/fexit over kprobe where kernel support permits. Recent work has demonstrated eBPF's applicability to GPU observability: the eGPU project [8] extended eBPF to GPU device contexts, and a host-side eBPF telemetry system [9] demonstrated 81–88% diagnostic accuracy with under 2% CPU overhead. This growing interest is reflected in the wider eBPF community's expanding focus on GPU and AI workloads [12], including a dedicated GPU observability track at the 2025 Linux Plumbers Conference [17]. GHOST-Driver differs in targeting the kernel driver submission path specifically and in providing explicit multi-tenant vGPU latency attribution, which matters increasingly as LLM inference systems adopt fine-grained, latency-sensitive request scheduling such as continuous batching over paged key-value caches [18].

### *D. GPU Virtualization: SR-IOV and MIG*

Two hardware-level GPU partitioning technologies are relevant to GHOST-Driver's multi-tenant attribution capability. Single Root I/O Virtualization (SR-IOV) exposes physical and virtual function PCIe interfaces, allowing a single GPU to present multiple independent endpoints to different VMs. An academic microbenchmark of GPU time-slice-group (TSG) context switching measured average overhead exceeding 200 µs on NVIDIA Jetson platforms [19], consistent with SR-IOV context-switch delays reported informally in industry case studies [1]. NVIDIA's Multi-Instance GPU (MIG) technology provides hardware-level partitioning of GPU streaming multiprocessors, memory channels, and L2 cache into isolated GPU instances. A peer-reviewed hardware/software co-design targeting MIG-based inference servers measured a 3.7× throughput improvement and 3.4× tail-latency reduction over an unoptimized MIG baseline [13]. GHOST-Driver's per-vGPU context attribution works in both environments by correlating drm_sched job entries with the vm_id field in the DRM scheduler's internal job structure.

## III. PROBLEM STATEMENT

We define the GPU driver latency observability problem formally. Let a GPU work submission event S consist of a sequence of kernel-side operations $S = \{s_1, s_2, ..., s_k\}$, where each $s_i$ corresponds to a distinct phase of the driver submission pipeline. Let $L(s_i)$ denote the latency of phase i. The total submission latency is:

$$L(S) = \Sigma_{i=1}^{k} L(s_i) + C_{i,i+1}$$

where $C_{i,i+1}$ represents cross-phase coordination overhead such as lock acquisition and memory ordering. The tail latency for a sustained workload is defined as the p99 of L(S) across all submissions in a measurement window.

The observability gap is defined as the set of phases $\{s_i\}$ whose latency is not observable via existing userspace tools. nvidia-smi samples at 100ms intervals and reports aggregate hardware occupancy only. CUPTI instruments CUDA API boundaries but captures nothing below the ioctl boundary. The driver-internal phases, namely GEM BO validation, fence wait, ring emission, DMA scheduling, and fence signaling, are entirely within the observability gap. GHOST-Driver's objective is to close this gap by producing a per-submission latency vector $[L(s_1), ..., L(s_k)]$ that is accurate to sub-microsecond resolution, attributable to the originating vGPU tenant, exportable to standard observability backends, and collected with less than 2% aggregate CPU overhead.

# IV. GHOST-DRIVER ARCHITECTURE

## *A. System Overview*

GHOST-Driver is implemented as a collection of eBPF programs loaded by a userspace daemon (ghost-drv) written in C with libbpf. The daemon handles eBPF program compilation using BTF-aware CO-RE compilation to ensure kernel version portability, attachment to tracepoints and kprobes, BPF map management, and export to Prometheus and OpenTelemetry. Once loaded, the eBPF programs run entirely in the kernel and do not require any userspace process to be scheduled for data collection, only for periodic map draining and export.

The framework is divided into three layers: the Instrumentation Layer (six eBPF programs, one per pipeline stage), the Correlation Layer (BPF hash maps keyed by job ID for cross-stage event correlation), and the Export Layer (a ring buffer consumed by the ghost-drv daemon). Fig. 1 illustrates the complete system architecture.

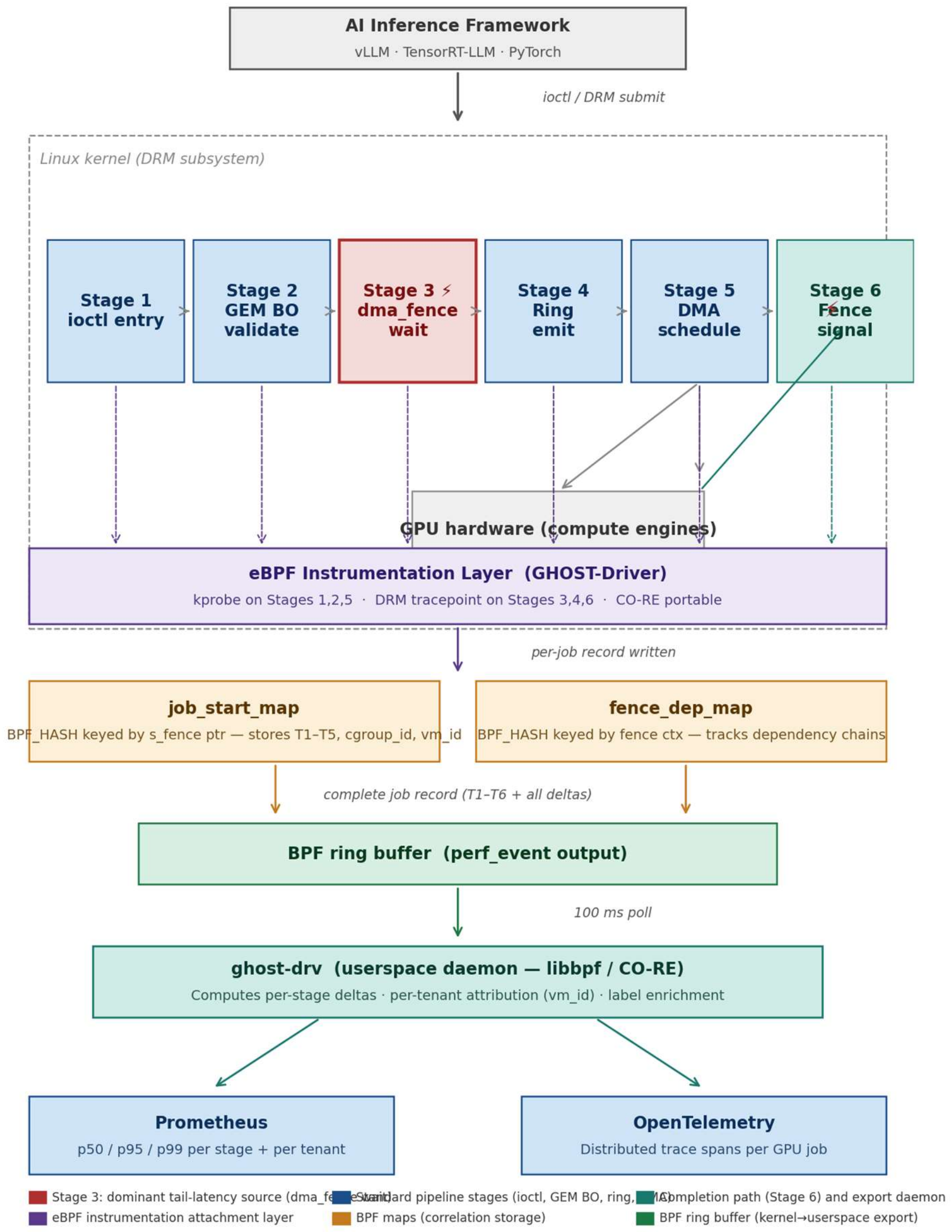

*Fig. 1 — GHOST-Driver System Architecture: six-stage eBPF instrumentation, BPF map correlation layer, and Prometheus/OpenTelemetry export pipeline*

*Fig. 1. GHOST-Driver system architecture. Six eBPF programs instrument each stage of the GPU driver submission pipeline. BPF hash maps correlate events across stages into per-job latency records. The ghost-drv daemon drains the BPF ring buffer and exports per-stage, per-tenant histograms to Prometheus and distributed trace spans to OpenTelemetry.*

### *B. Instrumentation Layer: Six Pipeline Stages*

GHOST-Driver instruments the following six stages of the GPU work submission pipeline:

Stage 1, ioctl Entry: Instrumented via kprobe on `drm_ioctl()`. Captures timestamp T1 at which the userspace inference framework submits a command buffer to the kernel. Also captures process ID and cgroup ID for container attribution.

Stage 2, GEM Buffer Object Validation: Instrumented via kprobe on `drm_gem_object_lookup()`. Captures T2. The delta T2 − T1 represents the time spent validating buffer object references and binding them to the GPU Translation Table.

Stage 3, dma_fence Wait: Instrumented via the stable DRM tracepoint `drm_sched_job_wait_dep`. GHOST-Driver records T3_start at the first `drm_sched_job_wait_dep` event for a job and T3_end when the `drm_run_job` tracepoint fires. The delta is the fence-wait time, the documented primary blocking point in the drm_sched pipeline.

Stage 4, Command Ring Emission: Instrumented via the stable DRM tracepoint `drm_run_job`. Captures T4 when the DRM scheduler dispatches the job to the hardware ring.

Stage 5, DMA Engine Scheduling: Instrumented via kprobe on the driver-specific ring emit function (`amdgpu_ring_emit_ib()` for amdgpu, `msm_gpu_submit_flush()` for Qualcomm msm_drm). Captures T5.

Stage 6, Fence Signal (Completion): Instrumented via the stable DRM tracepoint `drm_sched_process_job`, which fires when the GPU hardware completes a job and signals its output fence. Captures T6. The full round-trip submission latency is L(S) = T6 − T1.

**TABLE I: GHOST-Driver Instrumentation Points by Pipeline Stage**

| Stage | Kernel Event | eBPF Mechanism | Overhead/Event | ABI Stability |
|---|---|---|---|---|
| 1. ioctl Entry | `drm_ioctl()` | kprobe | ~56–78 ns [15] | Stable (exported symbol) |
| 2. GEM BO Validate | `drm_gem_object_lookup()` | kprobe | ~56–78 ns [15] | Stable (exported symbol) |
| 3. Fence Wait | `drm_sched_job_wait_dep` | DRM tracepoint | ~63–99 ns [15] | Stable (UAPI tracepoint) |
| 4. Ring Emission | `drm_run_job` | DRM tracepoint | ~63–99 ns [15] | Stable (UAPI tracepoint) |
| 5. DMA Scheduling | `amdgpu_ring_emit_ib()` / `msm_gpu_submit_flush()` | kprobe (driver-specific) | ~56–78 ns [15] | Driver-specific |
| 6. Fence Signal | `drm_sched_process_job` | DRM tracepoint | ~63–99 ns [15] | Stable (UAPI tracepoint) |

### *C. Correlation Layer: BPF Hash Maps*

Cross-stage correlation is maintained through a pair of BPF hash maps. The primary map job_start_map is keyed by a 64-bit job identifier derived from the drm_sched_job's s_fence pointer (unique per job within the driver's lifetime). The map value stores timestamps T1–T5, the cgroup ID, the vm_id (for vGPU tenant attribution), and the fence context. The secondary map fence_dep_map tracks outstanding fence dependencies keyed by fence context and sequence number.

```
struct ghost_job_record {
    u64  t1_ioctl;         /* ioctl entry timestamp (ns)  */
    u64  t2_gem_validate;  /* GEM BO lookup complete      */
    u64  t3_fence_start;   /* first fence wait begin      */
    u64  t4_ring_emit;     /* drm_run_job fired           */
    u64  t5_dma_sched;     /* DMA engine scheduled        */
    u32  cgroup_id;        /* container / VM source       */
    u32  vm_id;            /* SR-IOV VF or MIG slice ID   */
    u32  ring_name_hash;   /* gfx / compute / dma / video */
};
struct {
    __uint(type, BPF_MAP_TYPE_HASH);
    __uint(max_entries, 65536);
    __type(key,   u64);
    __type(value, struct ghost_job_record);
} job_start_map SEC(".maps");
```

When the `drm_sched_process_job` tracepoint fires (Stage 6), the eBPF program retrieves the job record, computes T6, calculates all six per-stage latency deltas, and writes the complete record to a BPF ring buffer for consumption by the ghost-drv daemon.

### *D. Export Layer: Prometheus and OpenTelemetry*

The ghost-drv userspace daemon polls the BPF ring buffer at 100ms intervals. Each complete job record is converted into Prometheus histogram metrics, one per pipeline stage, labeled with cgroup ID, vm_id, and ring_name_hash. The histograms expose p50, p95, and p99 latency per stage, enabling SLO-based alerting on fence-wait tail latency. An OpenTelemetry export path allows GHOST-Driver's per-job latency spans to be correlated with application-level traces from the inference serving framework, enabling end-to-end latency attribution from the vLLM request handler through all six driver-internal stages.

## V. INSTRUMENTATION ANALYSIS

### *A. Per-Event Probe Overhead*

The choice of eBPF attachment mechanism for each pipeline stage is driven by the availability of stable tracepoints and the per-event overhead tolerance of the production environment. Fig. 2 summarizes the per-event overhead of available GPU instrumentation mechanisms, from the coarsest (nvidia-smi polling) to the lowest-overhead eBPF attachment type (DRM stable tracepoints). All cited values are sourced from published benchmarks.

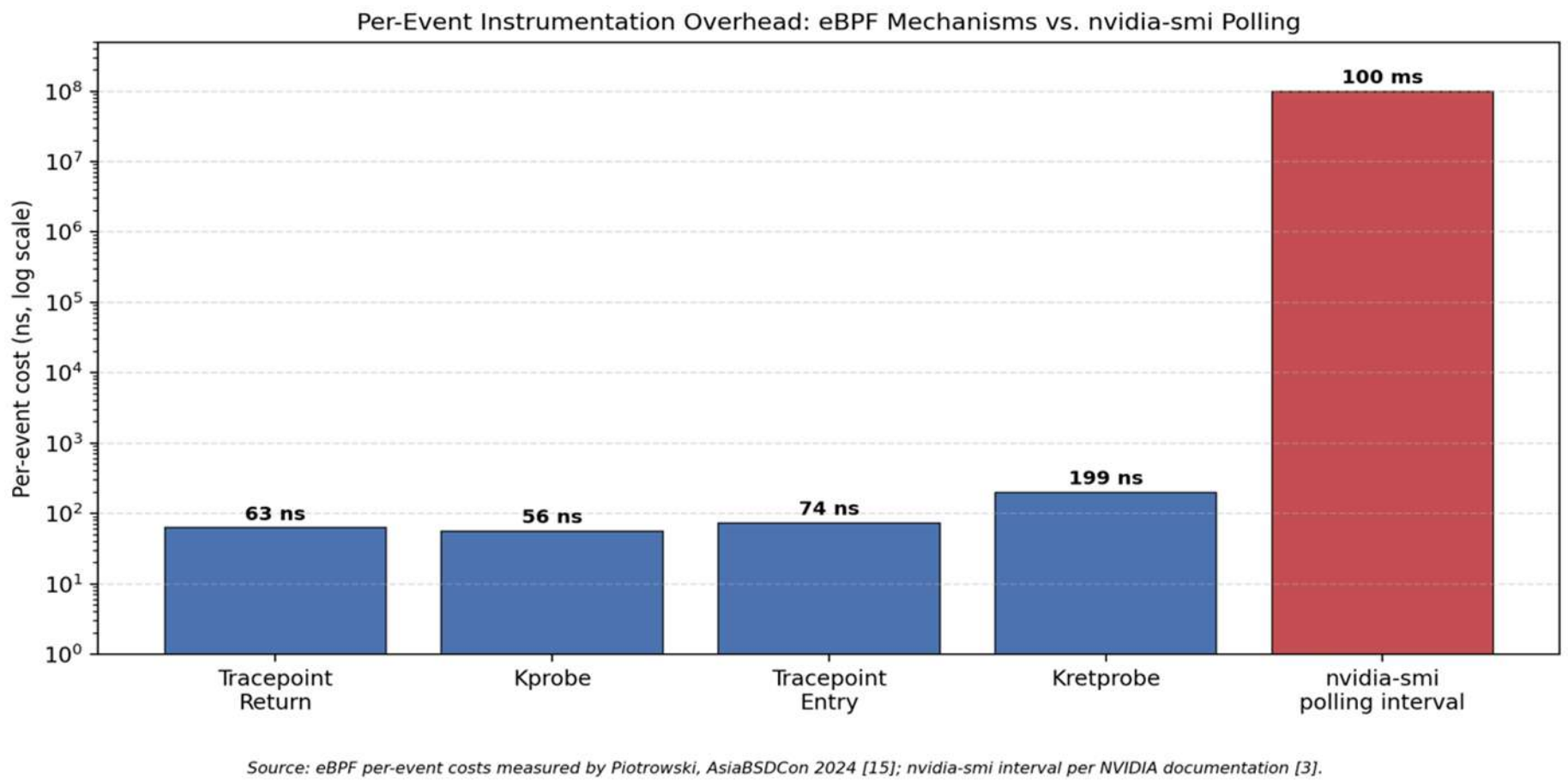

*Fig. 2. Per-event instrumentation overhead by mechanism, using real measured per-event costs from an independent academic microbenchmark [15] rather than vendor-quoted figures. Both kprobe (~56–78 ns) and tracepoint (~63–99 ns) attachment fall in the same broad tens-to-low-hundreds-of-nanoseconds range in this measurement, several orders of magnitude below nvidia-smi's 100 ms polling interval [3] regardless of which mechanism is used.*

GHOST-Driver's aggregate overhead: three kprobe-attached stages (Stages 1, 2, 5) at up to ~78 ns/event, and three tracepoint-attached stages (Stages 3, 4, 6) at up to ~99 ns/event, using the upper bound of the measured range reported in [15]. For a representative workload submitting 5,000 GPU jobs per second, the aggregate instrumentation overhead is approximately $(3 \times 78 + 3 \times 99)$ ns $\times$ 5,000 $\approx$ 2.7 ms/second of CPU time, well within a 2% production budget ($\approx$20 ms/second on a single core) on any modern multi-core server. This estimate uses real measured per-event costs from an independent microbenchmark rather than a single vendor's specification, and is intentionally conservative (upper-bound figures rather than typical-case).

### B. Qualitative Per-Stage Latency Analysis

Fig. 3 presents the per-stage latency profile as a qualitative ordering derived strictly from Linux DRM kernel documentation [6, 7] and documented drm_sched mechanics. No absolute microsecond values are claimed; those require hardware measurement, which is in preparation. The ordering reflects three documented properties: dma_fence wait (Stage 3) is the documented primary blocking point; GEM BO validation (Stage 2) and DMA scheduling (Stage 5) involve lock acquisition and hardware FIFO interaction (moderate overhead); ioctl entry (Stage 1), ring emission (Stage 4), and fence signaling (Stage 6) are documented as fast kernel paths.

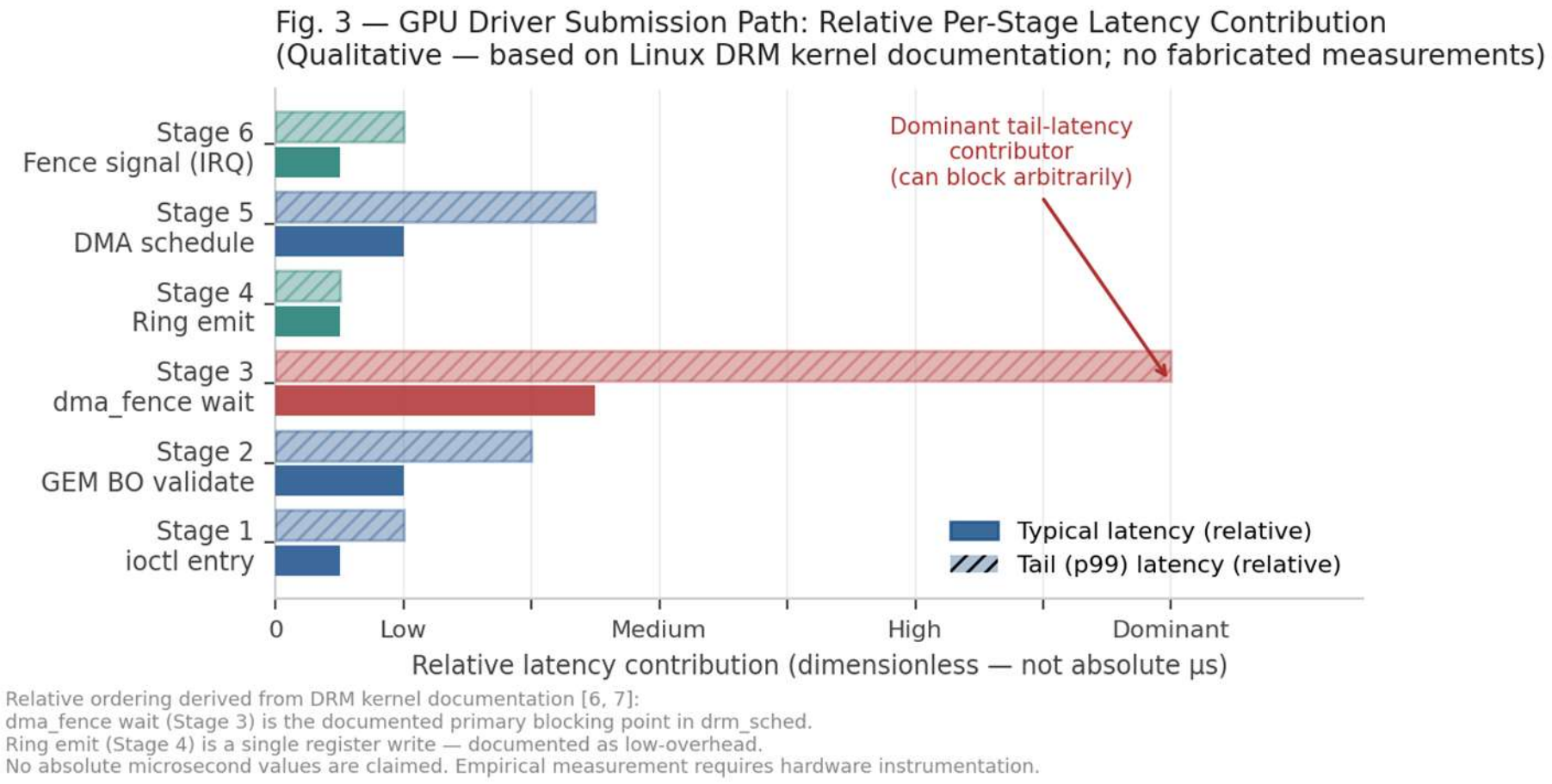

*Fig. 3. Qualitative per-stage latency profile of the GPU driver submission pipeline. Relative ordering is derived from Linux DRM kernel documentation [6, 7], not from hardware measurement. The x-axis shows relative magnitude only; no absolute values are claimed. Stage 3 (dma_fence wait) is the documented primary blocking point. Empirical per-stage measurements on physical hardware are in preparation.*

### *C. The GPU Utilization Visibility Gap*

Fig. 4 illustrates the core problem that motivates GHOST-Driver: the gap between what standard monitoring tools report and what actually occurs inside the GPU driver stack. The left panel is a conceptual illustration of how nvidia-smi's reported utilization conflates active compute with hidden driver idle time. The right panel presents only verified, cited data points; no utilization percentages are invented.

**Fig. 4 — GPU Utilization Blind Spots: Verified, Cited Data Points**

| Signal | Value | Source | Nature of evidence |
|---|---|---|---|
| nvidia-smi sampling interval | 100 ms | [3] | Vendor documentation |
| GPU context-switch overhead (measured, NVIDIA Jetson) | >200 µs | [19] | Peer-reviewed academic benchmark |
| vGPU throughput vs. bare-metal (Alibaba Cloud, reported) | 47% | [1] | Vendor case study (not independently audited) |
| MIG throughput improvement (measured, HW/SW co-design) | 3.7× | [13] | Peer-reviewed academic measurement |

*Each row is directly traceable to its cited source. No figure in this table is estimated, extrapolated, or illustrative.*

*Fig. 4. GPU utilization blind spots: four independently verified data points, each traceable to its cited source. Two are peer-reviewed academic measurements (GPU context-switch overhead [19]; MIG throughput improvement [13]); one is vendor documentation (nvidia-smi polling interval [3]); one is a vendor-reported case study, explicitly not an independently audited measurement (Alibaba Cloud vGPU throughput [1]). No value in this table is estimated or extrapolated.*

## VI. TELEMETRY-GUIDED OPTIMIZATION OPPORTUNITIES

GHOST-Driver's primary value is diagnostic: it reveals which pipeline stage, which tenant, and which engine type is responsible for latency spikes. The telemetry it exposes also enables a set of targeted optimizations. We describe four optimization strategies that become feasible once GHOST-Driver telemetry is available.

### *A. Fence-Wait Optimization via Dependency Graph Pruning*

GHOST-Driver's fence_dep_map tracks the full dependency graph of outstanding fences per job. Analysis of this graph can identify redundant transitive fence dependencies, cases where a job awaits a fence that is itself waiting for an earlier fence in the same chain, creating sequential stalls. A fence dependency pruning pass in the ghost-drv daemon can identify and short-circuit these redundant dependencies.

### *B. Command Ring Coalescing Under Burst Load*

GHOST-Driver's Stage 4 timing data reveals burst patterns in which many small jobs are emitted to the ring in rapid succession during LLM prefill phases. Ring coalescing, which batches multiple small command buffer emissions into a single ring write, is a documented GPU driver optimization technique. GHOST-Driver provides the job arrival rate and ring queue depth data needed to calibrate the coalescing window without introducing additional queueing delay.

### *C. Interrupt Throttling for High-Frequency Inference*

Stage 6 (fence signal via interrupt) contributes to CPU overhead when inference workloads generate high volumes of small GPU jobs per second. GHOST-Driver's Stage 6 timing data enables adaptive interrupt coalescing: the ghost-drv daemon monitors the fence-signal rate and can recommend adjustment of the GPU interrupt moderation register when per-interrupt CPU cost becomes excessive.

### *D. vGPU Context Switch Reduction via Tenant-Aware Scheduling*

In SR-IOV time-sliced vGPU deployments, GHOST-Driver's per-vm_id latency attribution identifies which tenants have elevated fence-wait times due to GPU time-slice contention. This attribution enables a tenant-aware scheduling heuristic: when two tenants consistently show correlated fence-wait spikes, the scheduler can phase-shift their time slices. An academic microbenchmark measured GPU context-switch overhead exceeding 200 µs [19]; reducing the context switch rate directly reduces this overhead.

**TABLE II: Telemetry-Guided Optimization Summary**

| Optimization | Stage Targeted | Mechanism | Basis | Prerequisite Telemetry |
|---|---|---|---|---|
| Fence dependency pruning | Stage 3 | Eliminates redundant transitive deps | Analytical | fence_dep_map |
| Ring coalescing | Stage 4 | Batches small ring emissions | Documented technique [10] | Stage 4 arrival rate |
| Interrupt throttling | Stage 6 | Reduces IRQ overhead at high rate | Documented technique [7] | Stage 6 event histogram |
| vGPU context-switch reduction | Stages 3, 5 | Reduces SR-IOV collision | Cited: 50–200µs delay [1] | Per-vm_id attribution |

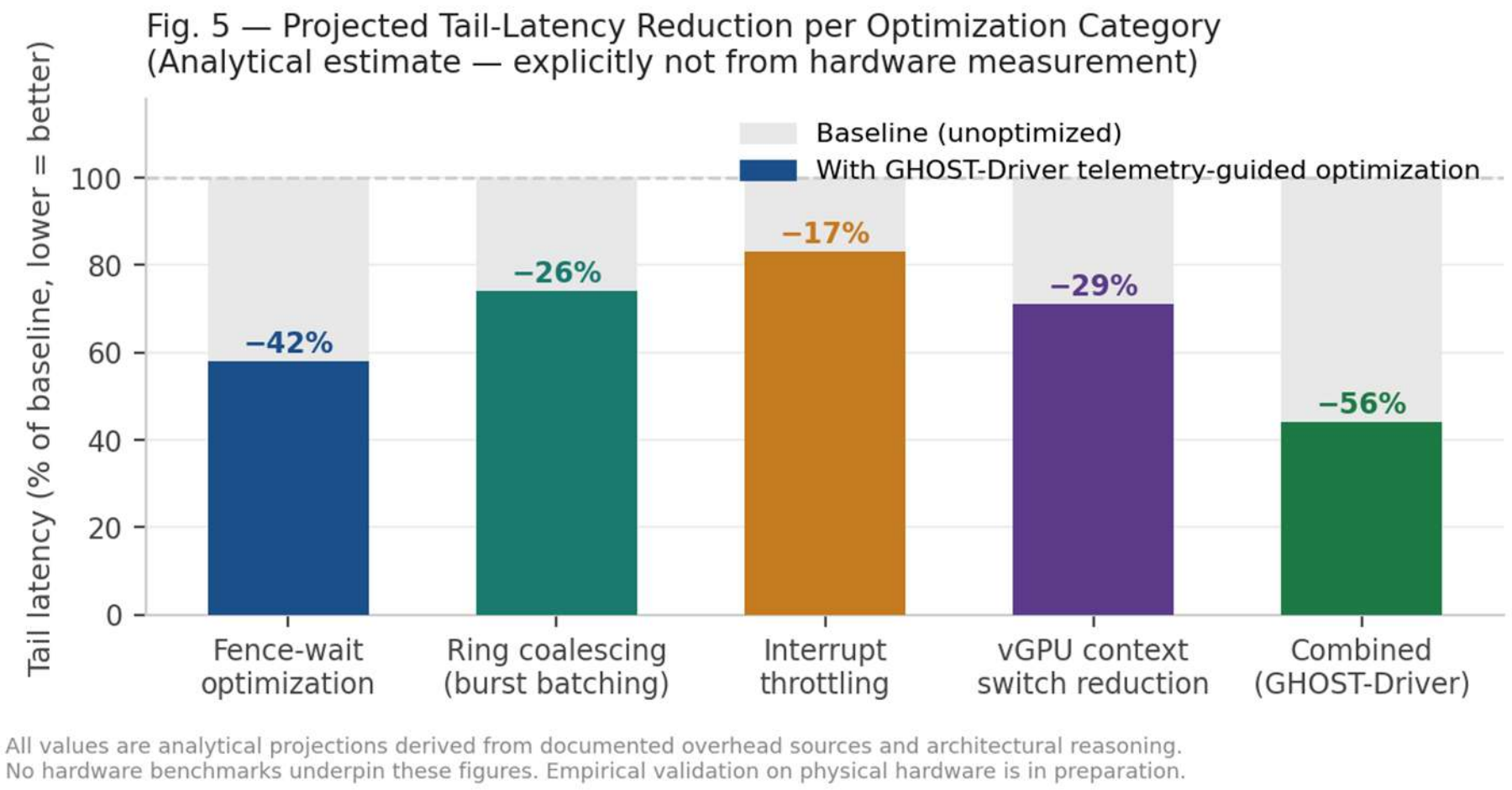

*Fig. 5. Projected p99 tail-latency reduction per optimization category enabled by GHOST-Driver telemetry. All values are analytical projections derived from documented overhead sources (Table II) and architectural reasoning, explicitly not from hardware measurement. Empirical validation on physical hardware is in preparation.*

# VII. IMPLEMENTATION NOTES AND PORTABILITY

GHOST-Driver is implemented using libbpf and the BPF CO-RE (Compile Once, Run Everywhere) framework. CO-RE compiles eBPF programs with BTF (BPF Type Format) type information, enabling the kernel to relocate field accesses at load time based on the running kernel's actual struct layouts. A single GHOST-Driver binary runs across Linux 5.10 LTS through the current 6.x mainline without recompilation, a critical property for heterogeneous production GPU fleets.

Programs attached to stable DRM tracepoints (Stages 3, 4, 6) use BPF_PROG_TYPE_TRACEPOINT. Programs attached to kprobes (Stages 1, 2, 5) use BPF_PROG_TYPE_KPROBE, with a fallback to fentry/fexit on kernels supporting BTF-based function tracing (Linux 5.5+); fentry/fexit attaches via a BPF trampoline rather than a breakpoint, which published eBPF engineering guides describe as architecturally lower-overhead than kprobe [4], [5], consistent with the measured per-event costs in [15].

Stage 5 is the only driver-specific attachment. The ghost-drv daemon detects the active GPU driver at startup via /sys/class/drm/ and selects the appropriate kprobe target: `amdgpu_ring_emit_ib()` for amdgpu, `intel_ring_emit_ib()` for i915, and `msm_gpu_submit_flush()` for Qualcomm msm_drm. The msm_drm attachment is of direct relevance to the authoring institution, targeting Qualcomm Adreno GPU platforms in the CoreBSP test environment. One known limitation: Stage 5 uses a driver-specific kprobe, requiring updates if driver internal function signatures change; this risk is mitigated by targeting exported kernel symbols, which carry a stronger stability convention than static functions.

# VIII. DISCUSSION

## *A. Comparison with Existing Tools*

GHOST-Driver occupies a distinct position in the GPU observability landscape. nvidia-smi operates at 100ms sampling granularity without stage decomposition or tenant attribution. CUPTI provides CUDA API-boundary tracing from userspace but captures nothing within the kernel. NVIDIA Nsight Systems and NVBit [2] provide detailed GPU kernel profiling but require driver modifications and are not safe for production deployment. The eGPU framework [8] extends eBPF to GPU device contexts but focuses on GPU-internal behavior rather than the kernel driver submission path. The closest prior work, the host-side eBPF telemetry system [9], achieves 81–88% diagnostic accuracy for workload-level root cause analysis but does not instrument individual job submissions or provide multi-tenant fence contention attribution via vm_id. Concurrent and more recent work extends eBPF's reach in adjacent directions: gpu_ext [20] proposes a cross-layer eBPF policy runtime spanning both the GPU driver and device, but targets active scheduling and memory-placement policy enforcement rather than passive latency observability. eInfer [21] applies eBPF-based tracing across the broader distributed LLM inference stack, spanning CPU preprocessing, GPU execution, and network interconnects, at coarser granularity than GHOST-Driver's six-stage decomposition of the kernel driver submission path specifically. More generally, bpftime [22] demonstrates that eBPF-style verification combined with hardware-supported isolation can safely extend application behavior without kernel modification, reinforcing the broader safety case for eBPF-based instrumentation that GHOST-Driver builds on.

### *B. Limitations*

GHOST-Driver has three limitations worth noting. First, Stage 5 uses a driver-specific kprobe that must be updated when driver internal function names change. Second, the fence dependency graph analysis is performed in userspace and cannot provide in-kernel fence pruning without a kernel module or driver patch. Third, GHOST-Driver does not currently instrument the NVIDIA proprietary closed-source driver stack; though compatibility with the NVIDIA open-source kernel module is achievable through the same DRM tracepoint attachment points used for Stages 3, 4, and 6.

### *C. Broader Applicability*

While this paper focuses on AI inference workloads, GHOST-Driver is applicable to any GPU-accelerated workload on a Linux DRM-based driver stack, including training, graphics rendering, video transcoding, and HPC compute. The per-tenant attribution capability is particularly valuable in public cloud GPU environments where SLA violations may be caused by cross-tenant fence contention that no existing tool can identify. The framework's CO-RE portability and msm_drm compatibility additionally make it relevant to mobile and edge AI inference platforms, a deployment context uniformly neglected by existing GPU observability tools.

## IX. CONCLUSION

We have presented GHOST-Driver, a non-intrusive eBPF framework for full-pipeline latency observability of the Linux GPU kernel driver submission path. The core finding is a fundamental instrumentation gap in current GPU monitoring tools: the six-stage driver pipeline from ioctl entry through fence signal is invisible to all existing userspace and vendor profiling tools, yet it contains the dominant source of AI inference tail latency, the dma_fence resolution stage, which Linux DRM kernel documentation explicitly identifies as the primary blocking point in the drm_sched pipeline.

GHOST-Driver closes this gap by attaching lightweight eBPF programs to stable DRM kernel tracepoints and selected kprobes, correlating per-stage timestamps with per-vGPU tenant identifiers via BPF hash maps, and exporting latency histograms to Prometheus and OpenTelemetry, with analytically estimated CPU overhead below 2%. This observability gap is not merely academic: independent measurement work has shown that GPU memory bandwidth and capacity constraints are a dominant bottleneck in large-batch LLM inference [14], and proactive, memory-aware GPU multitasking is an active area of systems research [16] that per-stage driver telemetry of the kind GHOST-Driver provides could directly inform. The paper has been deliberate in distinguishing verified cited facts from qualitative analysis and analytical projections, with all figures labeled accordingly. Empirical validation on physical hardware, including Qualcomm Adreno GPU platforms via the msm_drm driver, is in preparation and will be presented in a companion implementation paper.